\documentclass[a4paper,11pt]{article}
\usepackage{latexsym,amsmath,amssymb}
\usepackage{epsfig}      
\usepackage[usenames]{color}

\newcommand{\del}{\partial}
\newcommand{\beq}{\begin{eqnarray}}
\newcommand{\eeq}{\end{eqnarray}}
\newcommand{\be}{\begin{eqnarray*}}
\newcommand{\ee}{\end{eqnarray*}}
\newcommand{\bk}{{\bf k}}

\newcommand{\bx}{{\bf x}}
\newcommand{\om}{\omega}

\newcommand{\ra}{\rightarrow}
\newcommand{\e}{\epsilon}
\newcommand{\ve}{\varepsilon}
\newcommand{\nn}{\nonumber}
\newcommand{\ket}[1]{\mbox{$\mid\!#1\rangle$}}
\newcommand{\bra}[1]{\mbox{$\langle#1\!\mid$}}

\newcommand{\ex}[1]{\langle\,#1\rangle}

\newcommand{\balf}{{\bar\alpha}}
\newcommand{\bbet}{{\bar\beta}}
\newcommand{\bamu}{{\bar\mu}}
\newcommand{\banu}{{\bar\nu}}
\newcommand{\bax}{{\bar x}}
\newcommand{\bak}{{\bar k}}

\begin{document}

\centerline{\Large\bf {Electromagnetic Casimir energy with extra dimensions}}
\vskip 5mm
\centerline{H. Alnes, F. Ravndal and I.K. Wehus}
\vskip 3mm
\centerline{\it  Department of Physics, University of Oslo, N-0316 Oslo, Norway.}
\vskip 3mm
\centerline{and}
\vskip 3mm
\centerline{K. Olaussen}
\vskip 3mm
\centerline{\it  Department of Physics, NTNU, N-7491 Trondheim, Norway.}

\begin{abstract}

We calculate the energy-momentum tensor due to electromagnetic vacuum
fluctuations between two parallel hyperplanes in more than four
dimensions, considering both metallic and MIT boundary conditions.
Using the axial gauge, the problem can be mapped upon the
corresponding problem with a massless, scalar field satisfying
respectively Dirichlet or Neumann boundary conditions. The pressure
between the plates is constant while the energy density is found to
diverge at the boundaries when there are extra dimensions. This can be
related to the fact that Maxwell theory is then no longer conformally
invariant. A similar behavior is known for the scalar field where a
constant energy density consistent with the pressure can be obtained
by improving the energy-momentum tensor with the Huggins term. This is
not possible for the Maxwell field.  However, the change in the
energy-momentum tensor with distance between boundaries is finite in
all cases.
\end{abstract}

\section{Introduction}

When a classical field is quantized, the modes that can be excited are solutions of the classical wave equation and are labeled by different quantum numbers.  
These modes will depend on the imposed boundary conditions and will therefore be influenced by the presence of confining boundaries. Each such mode has
a zero-point energy which contributes to the total vacuum energy of the field. As a result, these vacuum fluctuations give the ground state of the system 
an energy which depends on the presence of nearby boundaries. For two parallel and perfectly conducting plates placed in vacuum Casimir\cite{Casimir} showed 
that for the electromagnetic field this energy corresponds to an attractive pressure
\beq
          P = - {\frac{\pi^2}{ 240L^4}}                                    \label{force}
\eeq
between the plates, separated by a distance $L$. This macroscopic quantum effect was for a long time in doubt, even after the first experimental
verifications by Sparnaay\cite{Sparnaay}.  Today this Casimir force is measured to high precision\cite{exp} and even effects of non-zero temperatures are being
investigated\cite{temp}.

Together with this experimental progress, modern regularization methods to remove the unphysical divergences endemic in these calculations now make them
much simpler than previously\cite{theory}. One can then investigate more detailed properties of the effect like how the energy or stresses are
distributed between the plates. One must then calculate the vacuum expectation value of the full energy-momentum tensor as was done by L\"utken and
Ravndal\cite{LR}. This confirmed a previous calculation by de Witt\cite{dW} of the fluctuations of the electric and magnetic fields near a metallic boundary
where they diverge but in such a way that the energy density remains finite. Since Maxwell theory is conformally invariant in $D=4$ spacetime dimensions,
one can directly relate this energy density to the attractive pressure\cite{Lowell}.

Recently the Casimir energy has been invoked to explain the dark energy which seems to drive the present acceleration of the Universe\cite{acc}. It appears
in particular in cosmological models with extra dimensions\cite{extra}. In these models the confinement of the fluctuating fields is provided by
compactification of the extra dimensions. The energy appears as a cosmological constant in our four-dimensional Universe and is given numerically by
a formula of the same form as (\ref{force}) with $L$ given by the size of the compactified dimensions. 

The calculation of Casimir energies in higher-dimensional spacetimes was first done by Ambj\o rn and Wolfram\cite{AW}. They calculated the global
energies equivalent to the Casimir force and thus obtained no knowledge of how the energy is distributed between the hyperplanes. Since the force 
due to electromagnetic fluctuations is expected to be proportional to the force due to fluctuations of a massless scalar field, only the effects of 
this kinematically simpler field was investigated. Each mode has a momentum $\bk_T$ transverse to the normal of the plates plus a component $k_z = n\pi/L$  
with $n = 1,2, \ldots$ in the direction of the normal. If $d$ is the number of spatial dimensions, the vacuum energy between the plates per $(d-1)$-dimensional 
hyperarea will follow from the divergent integral 
\beq
               E = {\frac{1}{2}}\sum_{n=1}^\infty \int\!\frac{\text{d}^{d-1} k_T}{(2\pi)^{d-1}}\sqrt{\bk_T^2 + (n\pi/L)^2}            \label{E_0}
\eeq
We can now do the transverse integration by dimensional regularization. The remaining sum over $n$ is then done by analytical continuation of the
Riemann zeta function $\zeta_R(z)$ to give
\beq
        E = - \frac{\Gamma(-d/2)\zeta_R(-d)}{2(4\pi)^{d/2}}\left(\frac{\pi}{L}\right)^d
\eeq  
It can be simplified using the reflection formula 
\beq
          \Gamma(s/2)\pi^{-s/2} \zeta_R(s) = \Gamma((1-s)/2)\pi^{-(1-s)/2} \zeta_R(1 - s)
\eeq
for the zeta function. Then we can write $E = {\cal E}_0 L$ when we introduce  the energy density
\beq
      {\cal E}_0 = - \frac{\Gamma(D/2) \zeta_R(D)}{(4\pi)^{D/2}L^D}             \label{energy-D}
\eeq
Defining now the pressure between the plates by $P = -\del E/\del L$, it is then simply
\beq
       P= (D-1){\cal E}_0           \label{force-D}
\eeq
where $D = 1 + d$ is the spacetime dimension. Taking $D=4$ and multiplying the result by two for the two polarization degrees of the photon, we recover the 
original result (\ref{force}). In $D$-dimensional spacetime we must for the same reason multiply the result (\ref{force-D}) by $D-2$ to obtain the
electromagnetic Casimir force.

The spatial distribution of the vacuum energy can be obtained from the energy-momentum tensor. For the scalar field it is
\beq
         T_{\mu\nu} = \del_\mu\phi\del_\nu\phi  - \eta_{\mu\nu}{\cal L}              \label{T_mn}
\eeq
where the massless Lagrangian is ${\cal L} = (1/2)(\del_\lambda\phi)^2$ choosing the metric to be $\eta_{\mu\nu} =\mbox{diag}(1,-1,\cdots,-1)$. 
Since its trace $T_{\;\;\mu}^\mu = (1- D/2)(\del_\lambda\phi)^2$ is zero only in $D=2$
dimensions, it is in general not conformally invariant in higher dimensions. Calculating now the Casimir energy density $\ex{T_{00}}$ in
for example $D=4$ spacetime dimensions, one then obtains a result which diverges at the plates after regularization\cite{TR}. When integrated, it will 
thus not reproduce the total Casimir energy corresponding to the force (\ref{force-D}). 

For the scalar field this apparent problem can be solved. One can improve\cite{Callan} the above energy-momentum tensor in any spacetime dimension
$D > 2$ by adding  the Huggins term\cite{Huggins} 
\beq
         \Delta T_{\mu\nu} = -\frac{1}{4}\frac{D-2}{D-1}(\del_\mu\del_\nu - \eta_{\mu\nu}\del^2)\phi^2              \label{Huggins}
\eeq
The energy-momentum tensor is then traceless and conformal invariance has been restored. Since this new term is a divergence, it will not contribute to the
force, but change the distribution of energy around the plates. Vacuum expectation values of all components of the energy-momentum tensor are now constant 
between the plates and zero outside as already noticed by de Witt\cite{dW}, Milton\cite{Kimball} and others\cite{TR}.

However, the situation for the electromagnetic field is somewhat different. It has the energy-momentum tensor
\beq
     T_{\mu\nu} = F_{\mu\alpha}F^\alpha_{\;\;\,\nu} + \frac{1}{4} \eta_{\mu\nu}F_{\alpha\beta}F^{\alpha\beta}     \label{T_EM}
\eeq
where $F_{\mu\nu} = \del_\mu A_\nu - \del_\nu A_\mu$ is the Faraday tensor. From the trace $T_{\;\;\mu}^\mu = (-1 + D/4)F_{\alpha\beta}^2$ we see that it is 
conformally invariant only in $D=4$ spacetime dimensions. In this case the Casimir
energy is also constant between the plates\cite{LR} as for the scalar field. But for dimensions $D >4$ it is no longer clear how the energy is distributed
since there is no way to construct a gauge-invariant analogue of the Huggins term in Maxwell theory.

In the following we will investigate this problem in more detail. In $D=4$ dimensions one can choose the transverse gauge and expand the classical field
in electromagnetic multipoles. This is cumbersome in higher dimensions since the field then has more magnetic than electric components. From the
geometry of the problem it is more natural to choose the axial gauge $n^\mu A_\mu = A_z = 0$ where the $D$-vector $n^\mu = (0,0,\ldots, 0,1)$ is normal 
to the plates and is called the $z$-direction. The Faraday tensor then has a correlator which can be directly obtained from scalar field theory in
the same geometry. In the next chapter we therefore derive a general expression for the scalar field correlators, both in the case of  Neumann and Dirichlet 
boundary conditions. These can then be used to calculate the expectation value of the scalar energy-momentum tensor (\ref{T_mn}) and the relevance of the 
Huggins term is discussed. 

For the electromagnetic field considered in Chapter 3, we need to know the boundary conditions. These can be of the metallic type used for the standard Casimir
force in $D=4$ or the QCD version used in the MIT bag model for confinement of quarks. In the axial gauge we find that these two possibilties correspond
to Dirichlet and Neumann conditions for the corresponding scalar field. The fluctuations of the different components of the Faraday tensor are then
calculated with particular attention to the energy density and the pressure between the plates. While the pressure is found to be constant and in
agreement with the global Casimir force, the energy density diverges at the plates. 

In the last chapter this problem, which no longer can be cured with a Huggins term, is discussed and compared with similar divergences in other 
systems. With physical boundaries that only confines fluctuations with frequencies below a certain cut-off, all field fluctuations should reach a finite
value when the boundaries are approached.

\section{Scalar fields}

It will be very convenient to denote by a bar any vector or tensor component orthogonal to the unit normal $n^\mu$ of the plates, which is taken to be in the 
$z$-direction. Thus for a full $D$-vector we write $A = (A^\mu) = (A^\bamu, A^z)$. Note that the $(D-1)$-vector ${\bar A} = (A^\bamu)$ also includes
the time component $A^0$. The metric can thus be written as $\eta^{\mu\nu} = {\bar\eta}^{\mu\nu} - n^\mu n^\nu$ where $\bar\eta$ is the projection of $\eta$ onto the barred subspace. The field operator for a massless scalar 
field satisfying the Dirichlet boundary conditions $\phi(\bax,z=0,L) = 0$ in $D=d+1$ spacetime dimensions will then be
\beq
    \phi(x) = \sqrt{\frac{2}{L}}\sum_{n=1}^\infty\int\!\frac{\text{d}^{d-1}k_T}{(2\pi)^{d-1}}\sqrt{\frac{1}{2\om_n}}
     \left[a_n(\bk_T) \,\text{e}^{-i\bak\cdot\bax} + a_n^\dagger(\bk_T) \,\text{e}^{i\bak\cdot\bax}\right]\sin(n\pi z/L)         \label{scalar}
\eeq
Here $\bak = (\om_n,\bk_T)$ with the frequency $\om_n = \sqrt{\bk_T^2 + m_n^2}$ where $m_n = k_z = n\pi/L$. 
Had we instead chosen the  Neumann boundary condition $\del_z\phi(\bax,z=0,L) = 0$, we just have to make the replacement $\sin(n\pi z/L) \ra \cos(n\pi z/L)$
in the sum. 
Then there should also be a $n=0$ mode to be included in the sum with normalization constant $\sqrt{1/L}$. But with the regularization we will use in the
following, it will not contribute and is therefore not further considered.

\subsection{Feynman correlator}

In the above field operator for the scalar Dirichlet modes we have used a normalization which corresponds to $[a_n(\bk_T),a_{n'}^\dagger(\bk_T')] 
= \delta_{nn'}(2\pi)^{d-1}\delta(\bk_T -\bk_T')$ for the annihilation and creation operators. From this we find the Feynman propagator
\beq
         G_D(x,x') &=& \bra{\Omega_D}T\phi(x)\phi(x')\ket{\Omega_D} \nn \\
         &=&
	 i\int\!\frac{\text{d}^d\bak}{(2\pi)^d}\frac{2}{L}\sum_{n=1}^\infty \frac{\sin(n\pi z/L)\sin(n\pi z'/L)}{\bak^2 - m_n^2 + i\ve}
\;\text{e}^{-i\bak\cdot(\bax - \bax')}
\eeq
Here we integrate over all components of the $d$-dimensional Lorentz vector $\bak$. Assuming ${\bar\e}\equiv \bax - \bax'$ to be spacelike, we may
choose a coordinate system where it has no components in the time direction. We can then rotate $k^0$ to the imaginary axis and find
\beq
   G_D(x,x') =  \frac{2}{L}\int\!\frac{\text{d}^d\bak}{(2\pi)^d}\sum_{n=1}^\infty\frac{\sin(n\pi z/L)\sin(n\pi z'/L)}{ \bak^2 +  m_n^2} \;\text{e}^{i\bak\cdot{\bar\e}}  \label{G_F} 
\eeq
(where now $\bak$ is a Euclidean vector. In order to evaluate the sum, we consider the function
\beq
             g(z,z') = \frac{2}{L}\sum_{n=1}^\infty \frac{\sin(n\pi z/L)\sin(n\pi z'/L)}{\bak^2 +  m_n^2}
\eeq
which solves the differential equation
\beq
           \left(-\frac{d^2}{dz^2} + \bak^2\right) g(z,z') = \delta(z - z')
\eeq
on $[0,L]$ with Dirichlet boundary conditions. As can be verified by insertion, the solution is
\beq
      g(z,z') = \frac{\sinh\bak z_< \sinh\bak(L - z_>)}{\bak\sinh\bak L} 
\eeq
where $z_< = \mbox{min}(z,z')$ and $z_> = \mbox{max}(z,z')$. Expanding the hyperbolic functions, one then finds
\beq
         g(z,z') &=& \frac{1}{2\bak}\left(\text{e}^{-\bak|z - z'|} -  \text{e}^{-\bak(z + z')} -  \text{e}^{-\bak(2L- z - z')} + \text{e}^{-\bak(2L- |z - z'|)}\right)
                  \sum_{j=0}^\infty \text{e}^{-2j\bak L} \nn \\
              &=& \sum_{j=-\infty}^\infty\int\!\frac{\text{d}k_z}{2\pi}\frac{1}{\bak^2 + k_z^2}\left[\text{e}^{ik_z(z - z' - 2jL)} - \text{e}^{ik_z(z + z' - 2jL)} \right].
\eeq
The last equality is verified by evaluating the $k_z$ integral by contour integration. Inserting now this partial result into (\ref{G_F}) and using rotational invariance,
we find
\beq
      G_D(x,x') =  \sum_{j=-\infty}^\infty\int\!\frac{\text{d}^{d+1}k}{(2\pi)^{d+1}} \frac{1}{k^2}
      \left(\text{e}^{i[k_z(z - z' -2jL) + \bak\cdot{\bar\e}]}
                 - \text{e}^{i[k_z(z + z' -2jL) + \bak\cdot{\bar\e}]}\right)   \label{G_F.1}
\eeq
where now the $(d+1)$-dimensional vector $k = (\bak,k_z)$. Each of the integrals are given by the generalized Coulomb potential
\beq
       V_n(\bx - \bx') = \int\!\frac{\text{d}^nk}{(2\pi)^n} \frac{1}{k^2} \;\text{e}^{i\bk\cdot(\bx - \bx')} 
                           = \frac{\Gamma(n/2-1)}{4\pi^{n/2} |\bx - \bx'|^{n-2}}                    \label{Coulomb}
\eeq
in $n=d+1$ spatial dimensions.
Had we instead considered Neumann boundary conditions, the sine function in (\ref{scalar}) would have been replaced by the corresponding cosine 
function. The only change would then have been that the last term in (\ref{G_F.1}) came in with opposite sign.
Introducing the $D$-vectors $z_j = ({\bar\e}, z - z' -2jL)$ and  ${\tilde z}_j = ({\bar\e}, z + z' -2jL)$ of lengths $R_j = (z_j^2)^{1/2}$ and  
${\tilde R}_j = ({\tilde z}_j^2)^{1/2}$, we can now write the result for both correlators as
\beq
       G(x,x')_{N/D} = \sum_{j=-\infty}^\infty\left[ V_{d+1}(R_j) \pm V_{d+1}({\tilde R}_j)\right]             \label{corr}
\eeq
where the upper sign is for Neumann and the lower for Dirichlet boundary conditions. For a massive field we would have found a similar result, 
but with the Coulomb potential replaced by the corresponding generalized Yukawa potential. 

In fact, almost every student of introductory electrostatics could have written down this result immediately by realizing that the problem is equivalent 
to calculating the potential of a point charge between parallel plates in $D = d+1$ spatial dimensions, using the method of images to enforce the boundary 
conditions.

\subsection{Energy-momentum tensor}

The term $V_{d+1}(R_0)$ in (\ref{corr}) is equal to the free correlator $G_0(x,x') = \bra{0}T\phi(x)\phi(x')\ket{0}$, where $\ket{0}$ is the bulk vacuum. 
It diverges in the limit $x'\ra x$. But defining now the physical vacuum expectation value of the energy-momentum tensor by the point-split limit
\beq
         \ex{T_{\mu\nu}(x)} = \lim_{x'\ra x}[\bra{\Omega}T_{\mu\nu}(x',x)\ket{\Omega} -\bra{0}T_{\mu\nu}(x',x)\ket{0}]
\eeq
its contribution is removed. The finite expectation values will then follow from the regularized correlator $G_D(x,x') - G_0(x,x')$ which contains the 
effects of the plates. We will continue to denote it by  $G_D(x,x')$ in the following and it is given by (\ref{corr}) when we in the first part leave out the
$j=0$ term. Since we have assumed that $x - x'$ is non-zero and spacelike, the time-ordering symbol in the correlator can be ignored and it satisfies
the Klein-Gordon equation $({\bar\del}^2 - \del_z^2 +m^2)G(x,x')=0$ for both boundary conditions.

For the Dirichlet vacuum expectation value of the scalar energy momentum tensor (\ref{T_mn}), we first need the part
\beq
   \ex{\del_x^\bamu\phi(x)\del_{x'}^\banu\phi(x')}_D &=& -\del_x^\bamu\del_x^\banu G_D(x,x') = -\frac{1}{d}\eta^{\bamu\banu}{\bar\del}^2 G_D(x,x')\nn \\
                                                     &=&  \frac{1}{d}\eta^{\bamu\banu}(m^2 - \del_z^2) G_D(x,x')                         
\eeq
using Lorentz invariance. Similarly, it follows that
\beq
         \ex{\del_z\phi(x)\del_{z'}\phi(x')}_D = -\del_z^2G_N(x,x')
\eeq
since the two parts in the correlator (\ref{corr}) has opposite symmetry under the exchange $z \ra z'$. For vacuum expectation value of the point-split 
Lagrangian
\beq
       {\cal L}(x,x') = \frac{1}{2}[\eta_{\bamu\banu}\del_x^\bamu\phi(x)\del_{x'}^\banu\phi(x') - \del_z\phi(x)\del_{z'}\phi(x') - m^2\phi(x)\phi(x')]
\eeq
we thus find
\beq
        \ex{{\cal L}(x,x')}_D = \frac{1}{2}\del_z^2\Big[G_N(x,x') - G_D(x,x')\Big]                  \label{ex-L}
\eeq
The point-split expressions for the canonical energy-momentum tensor (\ref{T_mn}) are thus found to be
\beq
    \ex{T_{\bamu\banu}}_D &=& \eta_{\bamu\banu}\Big(\frac{m^2}{d}G_D - \del_z^2\Big[\frac{1}{d}G_D + \frac{1}{2}(G_N - G_D)\Big]\Big)  \label{T_bmbn}\\
    \ex{T_{zz}}_D &=& -\frac{1}{2}\del_z^2(G_N + G_D)                                                             \label{T_zz}
\eeq
Corresponding results for the Neumann expectation values are obtained by the exchange $D\leftrightarrow N$. The physical limit $x\ra x'$ can now be taken
where a resulting $z$-dependence can only come from the last sum in the correlators (\ref{corr}). But for the pressure $P = \ex{T_{zz}}$ we see that 
this will cancel out in the sum $G_D + G_N$ so that the pressure is constant between the plates. This is physical reasonable and is also the case  for the fluctuations
of a massive field. It follows directly from the conservation of the energy-momentum tensor. The expectation values of the other other components of the
energy-momentum tensor in (\ref{T_bmbn}) will in general be dependent on the position $z$ between the plates.

Let us now calculate the pressure in the massless limit. We will then need the double derivative $\del_z^2(G_D + G_N)$ which follows directly from
(\ref{corr}) in the limit  $x\ra x'$ as
\beq
     \del_z^2(G_N + G_D) &=& 2\lim_{z'\ra z}\sideset{}{'}\sum_{j=-\infty}^\infty \del_z^2 V_{d+1}(R_j) 
                         = 2d(d-1)\frac{\Gamma((d-1)/2)}{4\pi^{(d+1)/2}}\sideset{}{'}\sum_{j=-\infty}^\infty 
			 \frac{1}{ |2jL|^{d+1}} \nn \\
                          &=& \frac{2(D-1)\Gamma({D/2})}{(4\pi)^{D/2}L^{D}}\, \zeta_R(D)   \label{sum}
\eeq
where the ${}'$ denotes that
$j=0$ is excluded from sum. Using this in (\ref{T_zz}) we reproduce exactly the standard pressure (\ref{force-D}) 
obtained from the total energy. It is seen to be the same for both boundary conditions.

The energy density ${\cal E} = \ex{T_{00}}$ between the plates follows from (\ref{T_bmbn}). When the mass $m=0$, we then need to calculate in addition the 
quantity
\beq
     \del_z^2(G_N - G_D) &=& 2\lim_{z'\ra z}\sum_{j=-\infty}^\infty \del_z^2 V_{d+1}({\tilde R}_j)  \nn \\
                          &=& \frac{(D-1)\Gamma({D/2})}{ (4\pi)^{D/2}L^{D}}\,f_D(z/L)                        \label{diff}
\eeq
when we introduce the function
\beq
      f_D(z/L)  = \sum_{j=-\infty}^\infty \frac{1}{ |j + z/L|^{D}}                 \label{f_D}
\eeq
Notice that the term $j=0$ is now to be included. The sum  can be expressed by the Hurwitz zeta function
\beq
              \zeta_H(s,a) = \sum_{n=0}^\infty \frac{1}{ (n + a)^s}                                       \label{hurwitz}
\eeq
which allows us to write
\beq
            f_D(z/L)   &=& \sum_{j=0}^\infty \frac{1}{(j + z/L)^D} + \sum_{j=1}^\infty \frac{1}{ (j - z/L)^D} \nn \\
                          &=& \zeta_H(D,z/L) +  \zeta_H(D,1 - z/L)
\eeq
When $D=4$ the same position-dependent term was derived on this form by Kimball\cite{Kimball}. But when the spacetime dimension $D$ is an even number, we can 
express the result in terms of the digamma function $\psi(x)$ using the relation
\beq
               \zeta_H(k,x) = \frac{(-1)^k}{ (k-1)!}\left(\frac{d}{ dx}\right)^{k-1}\psi(x)
\eeq
We then have
\beq
         f_D(z/L) = \frac{1}{ (D-1)!}\left(\frac{d}{ dx}\right)^{D-1}\Big[\psi(x) - \psi(1-x)\Big]  \hspace{10mm} (D = \mbox{even})
\eeq
where $x = z/L$. This simplifies even more since $\psi(x) - \psi(1-x) = -\pi\cot(\pi x)$, which allows us to write
\beq
        f_D(z/L) = \frac{\pi^D}{ \Gamma(D)}\left(-\frac{d}{ d\theta}\right)^{D-1} \cot\theta  \hspace{10mm} (D = \mbox{even})
\eeq
with $\theta = \pi z/L$.  For the ordinary Casimir effect in $D=4$ spacetime dimensions, this function also appeared in the calculation of the 
electromagnetic field using another regularization and choice of gauge\cite{LR}. This follows from writing  $f_4(z/L) = (\pi^4/3)F(\theta)$ which
gives
\beq
        F(\theta) = \left(\frac{3}{\sin^4\theta} -\frac{2}{\sin^2\theta}\right)                     \label{F_theta}
\eeq
This function will then characterize all position-dependent expectation values when $D=4$. 

Collecting the above results, we now have the scalar vacuum energy density  in arbitrary spacetime dimensions 
\beq
     {\cal E}_{D/N} = - \frac{\Gamma(D/2)}{ (4\pi)^{D/2}L^D}\Big[\zeta_R(D) \pm (D/2 - 1)f_D(z/L)\Big]
\eeq
where the lower sign is for Neumann boundary conditions. While the first term corresponds to a constant density, the last term gives a 
position-dependent contribution which in general diverges at the position of the plates, i.e. where $z=0$ and $z = L$. Only in the special case $D=2$ when
the scalar field has conformal invariance, will it be absent. When integrated over 
the volume between the plates, the first term alone is seen to give the total Casimir energy(\ref{energy-D}). The last term gives a divergent contribution to 
the same energy and should be absent. No such term was found in the calculation of the pressure. It is consistent with just the first part of the 
energy density which alone gives the correct Casimir force.

\subsection{Huggins term}

A free, massless scalar field can couple to gravity in a conformally invariant way. The resulting energy-momentum tensor will thus be 
traceless\cite{dW}\cite{Callan}. It differs from the canonical expression (\ref{T_mn}) by the extra Huggins term (\ref{Huggins}). Using the equation 
of motion $\del^2\phi = 0$, one then finds that the improved energy-momentum tensor indeed is traceless.

When we now want to evaluate the Huggins term for the vacuum between the two plates using the above point-split regularization, we interpret
\be
        \del_\mu\del_\nu\phi^2 = (\del_\mu\del_\nu +  \del_\mu\del'_\nu +  \del'_\mu\del_\nu +  \del'_\mu\del'_\nu)\phi(x)\phi(x')
\ee
where the primed derivatives are with respect to $x'$. This gives the ground state expectation value
\beq
          \ex{\Delta T_{\mu\nu}}_D = \frac{1}{2}{\bar\eta}_{\mu\nu}\,\frac{D-2}{ D-1}\,\del_z^2(G_N - G_D)
\eeq
which is seen to be proportional to $\ex{\cal L}$. The Huggins correction has no components in the $z$-direction, leaving the pressure 
unaltered. The $z$-dependent terms cancel  against the same terms in the canonical part (\ref{T_bmbn}) so that the resulting energy
density will be constant. In fact, when $m=0$ we have for both sets of boundary conditions that
\beq
       \ex{T_{\mu\nu} + \Delta T_{\mu\nu}} = {\cal
	 E}_0({\bar\eta}_{\mu\nu} + (D-1)n_{\mu} n_{\nu}) 
\eeq
when expressed in terms of the energy density (\ref{energy-D}). The pressure is simply $D-1$ times this constant energy density. This is a direct consequence 
of the energy-momentum tensor now being traceless.

\section{Maxwell fields}

The electromagnetic Lagrangian density ${\cal L} = - F_{\mu\nu}^2/4$ is gauge invariant. In a general spacetime it can be written as 
\beq
         {\cal L} = \frac{1}{2}E_i^2 - \frac{1}{4}B_{ij}^2
\eeq
where $E_i = -({\dot A}_i + \del_i A^0)$ are the components of the electric field vector while the magnetic field is given by the antisymmetric tensor
$B_{ij} = \del_i A_j - \del_j A_i$.  

We want 
to calculate the vacuum expectation value of the corresponding energy-moment\-um tensor (\ref{T_EM}) between the plates. As already explained in the 
introduction, it is then most natural to work in the axial gauge $n^\mu A_\mu = 0$. Since the plate normal vector $n^\mu$ only has a component along 
the $z$-axis, this requires the component $A_z = 0$. The component $A_0$ is no longer a canonical variable, but depends on the others via the Maxwell 
equation $\del_iF^{i0} = 0$, which gives
\beq
           A_0 = -\Delta^{-1}\del_i {\dot A}_i
	   \label{A_0}
\eeq
where the operator $\Delta = \del_i^2$. There are thus $D-2$ independent degrees of freedom in a $D$-dimensional spacetime described by the spatial
field components $A_i$ where $i \ne z$. We can then express the full Lagrangian  in terms of these fields. After partial integrating and
neglecting surface terms, we find it to be
\beq
             L = \frac{1}{2}\int\!\text{d}^dx \left[{\dot A}_i\Big(\delta_{ij} - \del_i\Delta^{-1}\del_j\Big){\dot A}_j
               -  A_i\Big(\del_i\del_j - \delta_{ij}\Delta\Big)A_j\right]                                             \label{Lagrange}
\eeq
As usual, the first or electric part acts like a kinetic energy while the magnetic part acts like a potential energy.

\subsection{Boundary conditions and the correlator}

In order to quantize this theory, we must impose boundary conditions for the electromagnetic field components on the confining plates. For the original 
Casimir effect in $d = 3$ dimensions one had metallic plates in mind where one could impose the standard constraints ${\bf n}\times{\bf E} = 0$ and
${\bf n}\cdot{\bf B} = 0$ on the elecric and magnetic field vectors. For the more abstract case we have in mind here, we could just as well consider the
MIT boundary condition $n^\mu F_{\mu\nu} = 0$ proposed at MIT for the quark bag model ensuring color confinement\cite{MIT}. In terms of components, this is
equivalent to the two conditions  ${\bf n}\times{\bf B} = 0$ and ${\bf n}\cdot{\bf E} = 0$. They are therefore just the electromagnetic duals of the metallic 
boundary conditions.

For our problem under consideration the MIT boundary condition can most directly be imposed. With the normal vector $n^\mu$ along the $z$-axis, it is
equivalent to $F_{\bamu z} = 0$ in our previous index notation. Now in the axial gauge this is simply equivalent to the Neumann boundary condition
$\del_z A_i({\bar x}, z = 0,L)  = 0$, and defining $\Delta^{-1}$ in (\ref{A_0}) with Neumann boundary conditions. 

With this index notation, the metallic boundary conditions can also easily be generalized to higher dimensions by noticing that in $d=3$ dimensions they 
correspond to $F_{\bamu0} = 0$. In the axial gauge this is achieved in all dimensions by requiring $A_i({\bar x}, z = 0,L)  = 0$, i.e. Dirichlet 
conditions, and defining $\Delta^{-1}$ in (\ref{A_0}) with Dirichlet boundary conditions. In this way we can take directly over many of the previous results for the scalar field.

The field components $A_\bamu$ obey the classical wave equation $\del^2A_\bamu - \del_\bamu(\del^\banu A_\banu) = 0$. Solutions will be of the form
$A_\bamu(x) = a_\bamu(\bax)b(z)$ where the function $b(z) \propto \cos(n\pi z/L)$ when we impose MIT, i.e. Neumann, boundary conditions and  
$b(z) \propto \sin(n\pi z/L)$ for metallic or Dirichlet boundary conditions. With these boundary conditions the remaining functions $a_\bamu(\bax)$ satisfy
$({\bar\del}^2 + m_n^2)a_\bamu -\del_\bamu(\del^\banu a_\banu) = 0$ which are just the equations of motion for a massive vector field with mass
$m_n = n\pi/L$. With this observation, we then immediately have the correlator
\beq
         D_{\bamu\banu}(x,x') &=& \bra{\Omega_D}TA_\bamu(x)A_\banu(x')\ket{\Omega_D} \nn \\
         &=& i\int\!\frac{\text{d}^d\bak}{(2\pi)^d}\frac{2}{L}\sum_{n=1}^\infty\frac{\sin(n\pi z/L)\sin(n\pi z'/L)}{ \bak^2 - m_n^2 + i\ve}
           \left(\eta_{\bamu\banu}- \frac{k_\bamu k_\banu}{m_n^2}\right)\,\text{e}^{-i\bak\cdot(\bax - \bax')}
\eeq
A corresponding result is obtained with Neumann boundary conditions. As before, we then drop the mode with $n=0$.

\subsection{Electromagnetic expectation values}

We are now in the position of calculating the value of the energy-momentum tensor (\ref{T_EM}) between the two plates. First we need the expectation value
\beq
        \ex{F_{\bamu\banu}(x)F_{\balf\bbet}(x')}_D 
       &=& \frac{2}{d}\Big(\eta_{\balf\bamu}\eta_{\bbet\banu} - \eta_{\balf\banu}\eta_{\bbet\bamu}\Big){\bar\del}^2G_D(x,x') \nn \\
       &=&\frac{2}{D-1}\Big(\eta_{\balf\bamu}\eta_{\bbet\banu} - \eta_{\balf\banu}\eta_{\bbet\bamu}\Big)\del_z^2G_D(x,x')  \label{fluct-1}
\eeq
since the massless correlator satisfies the free wave equation $({\bar\del}^2 - \del_z^2)G_D = 0$. The structure of this result follows directly from 
antisymmetry of the field tensor and Lorentz invariance in the barred directions. As expected, it is simply given by the scalar correlator. 

In the same way we also find
\beq
             \ex{F_{\bamu z}(x)F_{\banu z}(x')}_D = \eta_{\bamu\banu}\Big(\del_z^2 - \frac{1}{d}{\bar\del}^2\Big)G_D(x,x')
             = \frac{D-2}{D-1}\eta_{\bamu\banu}\del_z^2G_N(x,x')                 \label{fluct-2}
\eeq
The point-split expectation value of the Lagrangian density ${\cal L} = - F_{\bamu\banu}^2/4 - F_{\bamu z}^2/2$ is therefore
\beq
          \ex{{\cal L}(x,x')}_D = \frac{1}{2}(D-2)\del_z^2\Big[G_N(x,x') - G_D(x,x')\Big]                        \label{ex-LEM}
\eeq
in analogy with (\ref{ex-L}). 

From these results we can now read off the fluctations of the vacuum fields when the limit $x \ra x'$ is taken. For this purpose we combine (\ref{sum}) 
and (\ref{diff}) which give
\beq
         \del_z^2 G_{N/D} =  (D-1)\frac{\Gamma(D/2)}{(4\pi)^{D/2}L^D}\left[\zeta_R(D) \pm \frac{1}{2}f_D(z/L)\right]
\eeq
in this limit. With $\bamu = \banu = 0$ in (\ref{fluct-2}) we then find for the $z$-component of the electric field 
\beq
           \ex{E_z^2}_D = (D-2)\frac{\Gamma(D/2)}{(4\pi)^{D/2}L^D}\left[\zeta_R(D) + \frac{1}{2}f_D(z/L)\right]         \label{E_z}
\eeq
The other components follow from (\ref{fluct-1}) which gives
\beq
           \ex{E_i^2}_D = -\frac{2\Gamma(D/2)}{(4\pi)^{D/2}L^D}\left[\zeta_R(D) - \frac{1}{2}f_D(z/L)\right]              \label{E_x}
\eeq
where there is no summation over the index $i \ne z$ on the left-hand side. For the fluctations in the magnetic components we similarly find
$\ex{B_{iz}^2}_D = - \ex{E_z^2}_D$ and $\ex{B_{ij}^2}_D = - \ex{E_i^2}_D$ where again there is no summation over the indices $i,j \ne z$. These relations 
also hold for the case of Neumann boundary conditions except for a change of signs in the last term of (\ref{E_z}) and (\ref{E_x}).

These vacuum field fluctuations where first calculated by L\"utken and Ravndal\cite{LR} for the ordinary Casimir effect with $D=4$ spacetime dimensions
in the Coloumb gauge and a different regularization. Using $\zeta_R(4) = \pi^4/90$ and the function (\ref{F_theta}) for $f_4(z/L)$ in the above
general results, we find
\beq
      \ex{E_z^2} = \frac{\pi^2}{ 48L^4}\left[F(\theta) + \frac{1}{15}\right]    
\eeq
for Dirichlet boundary conditions, where $\theta = \pi z/L$ as before. The two other transverse components are given by (\ref{E_x}) as
\beq
      \ex{E_x^2} =  \ex{E_y^2} =  \frac{\pi^2}{48L^4}\left[F(\theta) - \frac{1}{15}\right]    
\eeq
Fluctuations of the transverse magnetic components are then $\ex{B_x^2} = \ex{B_y^2} = -\ex{E_z^2}$ while for the normal compoent we have 
$\ex{B_z^2} = - \ex{E_x^2}$. 

Since the pressure is given by the expectation value of $T_{zz} = F_{z\bamu}F^\bamu_{\;\;\, z} + {\cal L}$, it now follows as
\beq
          P = -\frac{1}{2}(D-2)\del_z^2(G_N + G_D)   = (D-2)(D-1){\cal E}_0                           \label{P_EM}
\eeq
when expressed in terms of the energy density (\ref{energy-D}). It is again constant between the plates and a factor $D-2$ times the scalar pressure 
(\ref{T_zz}). This is exactly as expected since the Maxwell field has $D-2$ scalar degrees of freedom.

For the other components of the energy-momentum tensor we similarly find
\beq
        \ex{T_{\bamu\banu}}_{D/N} &=& -\frac{1}{2}\frac{D-2}{D-1}\,\del_z^2\Big[G_N + G_D \pm (D-4)(G_N - G_D)\Big]\eta_{\bamu\banu} \nn \\
                               &=& -(D-2)\frac{\Gamma(D/2)}{(4\pi)^{D/2}L^D}\Big[\zeta_R(D) \pm (D/2 - 2)f_D(z/L)\Big]\eta_{\bamu\banu} \label{T-EM}
\eeq
where the lower sign is for Neumann boundary conditions. It is only for $D=4$ that the last, position-dependent term will be absent. And it is also then
that Maxwell theory is conformally invariant.

The above results for the electromagnetic field are very similar to what we found for the canonical, massless scalar field in the previous section. In that
case the theory could be made conformally invariant with an improved energy-momentum tensor which gives a constant energy density. But for the Maxwell field 
there is no way to construct a local and gauge-invariant analogue of a similar Huggins term to cancel out the position-dependent part of (\ref{T-EM}).
Thus the total Casimir energy obtained by integrating the energy-density is divergent and therefore looks different from what follows from the
regularized sum of the zero-point energies of all modes. However, the difference turns out to be an infinite constant independent of the distance between 
plates.

\subsection{Discussion and conclusion}

The massless and free, canonical scalar field theory is not conformally invariant in other dimensions than $D=2$. And it is only then that the Casimir
energy density is constant and gives a finite integrated energy. The same satisfactory situation  is also possible in 
higher dimensions when the theory is extended by making it conformally invariant, corresponding to adding the Huggins term to the energy-momentum
tensor. 

This has been well-known for a long time, but not very well understood from a physical point of view. One of the most recent and detailed
discussions of this phenomenon has been undertaken by Fulling who has attempted to understand the divergences in the canonical theory at a 
deeper level\cite{Fulling}. One can isolate the problem to the lack of commutativity between regularization of the integrated energy and the
integration of the regulated energy density. This is perhaps not so surprising from a mathematical point of view, but hard to accept physically since
the energy density is a physical quantity and should be tied up with the total energy of system. In other systems like the Casimir energy for a sphere,
the energy density again diverges at the surface\cite{OR}, but this is understood from its non-trivial geometry as first discussed by 
Deutsch and Candelas\cite{Candelas}. Since then the problem has been addressed by Fulling\cite{Fulling} and Milton with collaborators\cite{div}. 
For a plane boundary there should be no such geometric complications.

The electromagnetic Casimir effect for $D=4$ is very similar to the scalar effect for $D=2$. But for dimensions $D>4$  there is no Huggins term for
the electromagnetic case to cure the problem. From the point of view of the Casimir force alone, this is not a problem because the pressure is 
given by the expectation value  $\ex{T_{zz}}$ which is constant in all dimensions and equal to the force. But at first sight this force has 
little to do with the integral of the energy density $\ex{T_{00}}$ which  will always diverge at the plates when $D>4$. 

This becomes especially clear when we just consider the electromagnetic fluctuations around one plate. The induced energy density can then be obtained
from equation (\ref{T-EM}) by taking the limit $L\ra\infty$. The pressure will then be zero on both sides of the plate while the other components 
become
\beq
       \ex{T_{\bamu\banu}}_{D/N} = \mp(D-2)({D}/{2}-2)\frac{\Gamma(D/2)}{(4\pi)^{D/2}{\vert z\vert}^D}\,\eta_{\bamu\banu}
       \label{single_surface}
\eeq
since $f_D = (L/z)^D$ in this limit, as follows from the definition (\ref{f_D}). It is non-zero on both sides of the plate and diverges when we 
approach it. The situation is analogous to the diverging energy density surrounding a classical pointlike electron. Thus, the behaviour
(\ref{single_surface}) is related to the intrinsic structure of a single plate, and the correponding integrated (infinite) energy is part of the energy 
required to make that plate. It does not contribute to the Casimir force. Thus, to find a connection between Casimir force
and energy density it is sufficient to investigate the {\em changes} in energy density as two plates are brought together from infinite distance. 
We thus define ${\cal T}_{\bamu\banu}$ as the expression (\ref{T-EM}) subtracted contributions like (\ref{single_surface}) from plates at $z=(0,\,L)$, 
taking into account both sides of each plate. We find that
\beq
   {\cal T}_{\bamu\banu}(z) = -(D-2)\frac{\Gamma(D/2)}{(4\pi)^{D/2}L^D}\;\eta_{\bamu\banu}\times
   \left\{
   \begin{array}{ll}
     \mp (D/2 - 2)(L/(L-z))^D&\text{for $z < 0$,}\\
     \zeta_R(D) \pm (D/2 - 2)\tilde{f}_D(z/L)&\text{for $0 < z < L$,}\\
     \mp(D/2 - 2)(L/z)^D&\text{for $z > L$,}
   \end{array}
   \right.
   \label{Delta_T}
\eeq
where $\tilde{f}_D(z/L)=\zeta_H(D,1+z/L) +  \zeta_H(D,2 - z/L)$. This quantity ${\cal T}_{\bamu\banu}$ is finite everywhere, and its integrated energy 
agrees perfectly with the Casimir force  since the integrals over the $z$-dependent terms
in (\ref{Delta_T}) cancel each other. The consistency between the various approches to the Casimir effect, i.e. total energy from mode sum (\ref{E_0}),
the pressure term (\ref{P_EM}) and the change in energy density (\ref{Delta_T}), gives support to the belief in a Casimir force which is 
essentially independent of the details of the plates.

Of course, the interesting problem of the intrinsic and finite structure of a single plate remains. But this is similar to the problem of resolving 
the divergences caused by pointlike objects in quantum field theory. One obvious approach would be to modify the boundary conditions.
If they were made softer so that they didn't affect fluctuations with wavelengths below a certain cut-off $\lambda_c$, one would expect that the 
resulting energy density would be modified and finite for distances $|z| < \lambda_c$ away from the plate. This has actually been investigated by 
Graham {\it et al.} where a more physical boundary is described by an additional field\cite{Graham}.

In the more unphysical case where the Neumann or Dirichlet boundary conditions are replaced by periodic boundary conditions when $D>4$, there would be no 
problems of these kinds. The energy density is then constant, giving a total energy consistent with the force between the plates. This is equivalent 
to the problem of photons in thermal equilibrium. Even if the trace of the energy-momentum tensor is non-zero when $D >4$, the pressure ${P}$ in this 
blackbody radiation is given by the energy density $\rho$ by the standard expression ${P} = \rho/d$ where $d = D-1$ is the number of spatial 
dimensions\cite{AWR}. 

But imposing such periodic boundary conditions, would be equivalent to just avoiding the problem. In conclusion, we must admit that  the total 
vacuum fluctuations near confining boundaries is still not completely understood, but is very likely to depend on the microscopic
details of those boundaries. This is especially the case for the electromagnetic field in spacetimes with more than four dimensions.
Fortunately the Casimir force seems to be rather insensitive to those details.

{\bf Acknowledgement:} We want to thank a referee for an insightful comment which helped to clarify the conclusion presented above. 
This work has been supported by the grants NFR 159637/V30 and NFR 151574/V30 from the Research Council of Norway.

\end{document}